\magnification=1100
\baselineskip=12pt
\hsize = 13.4 true cm
\vsize = 22.2 true cm
\hoffset=1.0 true cm
\font\smallit = cmti8 at 8 pt
\input psfig.sty
\def\ref{\par\noindent\hangindent 1.5 true cm}
\vbox{\baselineskip 9 pt \smallit\noindent
Planetary Nebulae and Their Role in the Universe
\par\noindent IAU Symposium, Vol. 209, 2002
\par\noindent R. Sutherland, S. Kwok, M. Dopita, eds.}
\bigskip
\bigskip
{\bf PN and galactic chemical evolution}
\vskip 1.0 true cm
Walter J. Maciel and Roberto D. D. Costa
\smallskip
{\it IAG/USP - S\~ao Paulo SP, Brazil}
\bigskip
\hskip 0.05 true cm\vbox{\hsize = 12.5 true cm \noindent
{\bf Abstract.} Recent applications of PN to the study of 
galactic chemical evolution are reviewed, such as PN and 
stellar populations, abundance gradients, including
their space and time variations, determination of 
the He/H radial gradient and of the helium-to-metals 
enrichment ratio, and the [O/Fe] $\times$ [Fe/H] 
relation in the solar neighbourhood and in the 
galactic bulge.}
\vskip 1.0 true cm
\noindent
{\bf 1. \ \ Introduction}
\bigskip\noindent
Chemical evolution models must satisfy a series of 
observational constraints, such as the metallicity 
distribution, the age-metallicity relation, the 
existence of abundance gradients and their spatial 
and temporal variations.

Planetary nebulae have an important role in the 
establishment of some of these constraints. Basically, 
two classes of elements can be considered. First, elements 
such as He  and N, which are clearly affected by the stellar 
evolution. Second, most of the heavier elements, such 
as S and Ar, which are probably unaltered during the 
evolution of the PN progenitor star, so that the nebular 
abundances of these elements are essentially the same 
as the interstellar abundances at the time when the 
progenitor stars were formed.

In this paper, some recent results concerning the 
application of PN to the study of galactic chemical
evolution are reviewed, including PN abundances and 
stellar populations, abundance gradients and their 
variations, the He/H radial gradient, the helium-to-metals
enrichment ratio, and the determination of the 
[O/Fe] $\times$ [Fe/H] relation in the solar neighbourhood 
and in the galactic bulge. Some recent related reviews 
are Maciel (1997; 2000a).
\bigskip\noindent
{\bf 2. \ \ PN and Stellar Populations}
\bigskip\noindent
In the framework of the classification scheme originally 
proposed by Peimbert (1978), PN can be classified into 
five types: Type~I are disk objects with relatively massive 
progenitors; Type~II are disk objects with average mass 
progenitors; Type~III are thick disk objects, kinematically 
detached; Type~IV are halo objects, and Type~V are bulge 
nebulae. The main criteria used in this classification scheme 
are the nebular abundances, the average distance $z$ to the 
galactic plane, and the peculiar radial velocity $\Delta v$.
Morphological aspects have also some connection with the different
types as discussed by Stanghellini (this volume). 

Average values of the abundances are given by Maciel 
(2000a; see also Costa et al. 1996). From such data, it 
can be concluded that the He/H ratio increases in the disk 
along the sequence III--II--I, similarly to N/H, S/H, Ar/H 
and Cl/H. For O/H and Ne/H, there is an increase from  
Type~III to Type~II, but the average abundances of Type~I 
PN are not clearly higher than for Type~II, which may be 
partially due to ON cycling in the progenitor stars.

The abundances relative to oxygen of the elements Ne, S, Ar, 
and Cl can be considered as representative of the 
interstellar medium at the time of formation of these stars. 
Except for halo PN, all objects have essentially constant 
ratios, a result that has been confirmed for S, Ne and Ar 
for galactic and extragalactic HII regions 
(Henry \& Worthey 1999). 
\bigskip\noindent
{\bf 3. \ \ Abundance Gradients from PN}
\bigskip\noindent
The magnitude of the abundance gradients can be derived from 
a number of different sources, such as HII regions, PN, 
supernova remnants, and several types of stars, such as B stars, 
open cluster stars and cepheids (Henry \& Worthey 1999;
Smartt 2000). Average values are $d\log ({\rm O/H})/dR \simeq -0.06$ 
to $-0.07$ dex/kpc, with an uncertainty of about 0.02 dex/kpc.

Recent work on photoionized nebulae have established the 
presence of gradients for O/H and other elements such as N/H, 
S/H, Ne/H and Ar/H (see for example Maciel 1997; 2000a). 
Similar gradients are also observed in spiral 
galaxies (see for example Ferguson, Gallagher \& Wyse 1998). 

Stellar data based on O and B stars have been more controversial, 
but most recent determinations imply gradients of the same order 
as the one derived from photoionized nebulae (Rolleston et al. 
2000).  Data on cepheid variables should  in principle agree with 
the B star results, which is confirmed by a recent analysis by 
Caputo et al. (2001), but a flatter gradient and/or a discontinuity 
around 10 kpc is suggested from the high resolution data by 
Andrievsky et al. (2002). 

The most detailed work on abundance gradients from PN is that 
of Maciel \& Quireza (1999), who have increased and updated the 
sample by Maciel \& K\"oppen (1994) and studied gradients of the
ratios O/H, Ne/H, S/H and Ar/H. Results show that linear fits 
to the gradients are $-0.058$, $-0.036$, $-0.077$, and $-0.051$ 
dex/kpc for O/H, Ne/H, S/H and Ar/H, respectively. 

From the study of Maciel \& Quireza (1999), some evidence is 
found for a flattening of the gradients at large radial distances, 
especially for the O/H and S/H ratios. However, there are very 
few nebulae at distances $R > 12$ kpc, so that these results 
cannot be considered as definitive. Preliminary results of a
project to derive abundances for PN in the anticentre direction
(Costa, Maciel \& Uchida, this volume) suggest that the 
flattening is real. Some flattening in the observed gradients has
been obtained in some recent work on HII regions (cf. V\'\i lchez 
\& Esteban 1996), but regarding the stellar data, current results 
reported by Smartt (2000) apparently show no evidences of 
flattening up to 16 kpc from the galactic centre. 

The time variation of the abundance gradients is still poorly 
known. The best determined gradient from Type~II PN, which is 
that of O/H, is similar to the average value derived 
from the younger population of HII regions and hot stars,
within the uncertainties. Earlier work by Maciel \& K\"oppen
(1994) suggested that Type III PN show flatter gradients than
Type~II PN, but part of this may be due to orbital diffusion. 
On the basis of these results,  Maciel and Quireza (1999) 
have made a rough estimate of the average steepening rate 
as $-0.004$ dex kpc$^{-1}$ Gyr$^{-1}$.  An update of this 
estimate, taking into account HII regions, B stars and 
Type~II PN leads to a lower value, $-0.002$ dex kpc$^{-1}$ 
Gyr$^{-1}$. Although this result is clearly uncertain, it 
shows that any time variation in the gradients is probably 
small, so that the possibility of a constant gradient 
cannot be ruled out, at least for the last 5 Gyr, which 
approximately corresponds to the ages of the majority of 
the PN central stars. This agrees with the fact that 
most non-barred spiral galaxies have similar gradients, 
so that their gradients have probably not changed very 
much during their evolution. 

We have taken into account the PN sample of Maciel and K\"oppen 
(1994) and Maciel and Quireza (1999), supplemented by about 
40 PN near the anticentre direction from Costa et al. (this
volume). For these objects, we have estimated [Fe/H] 
metallicities and individual ages, according to the 
[O/H] $\times$ [Fe/H] and the age-metallicity relations given 
by Edvardsson et al. (1993). As a result, relatively 
accurate gradients can be derived for PN in a given age
group, which in principle allows for a better determination 
of the time evolution of the gradients as compared with 
previous estimates based entirely on the Peimbert 
types. In fact, there is some evidence for some mass -- and
therefore age -- overlapping for a given Peimbert type 
(see for example Peimbert and Carigi 1998), so that the 
determination of individual ages  probably gives a more 
accurate gradient variation. Preliminary results are shown 
in figure 1. In figure 1a, O/H gradients are shown for 
PN with ages under 3 Gyr (squares), between 3 and 6 Gyr
(solid dots) and higher than 6 Gyr (crosses). Figure 1b shows a 
similar plot, except that the age groups are: under 4 Gyr 
(squares), between 4 and 5 Gyr (solid dots) and higher than 5 Gyr
(crosses). Different definitions of the age groups are made 
in order to obtain groups of similar sizes. It can be seen 
that younger PN show flatter gradients. This tendency is clear 
in both figures, and is particularly strong for Groups III and 
II (figure 1a) and Groups II and I (figure 1b).  In figure 1a, 
the gradients have flattened from $-0.11\,$dex/kpc to
$-0.06\,$dex/kpc, while for figure 1b we have
$-0.09\,$dex/kpc for the oldest group and
$-0.05\,$dex/kpc for the youngest one. Overall, once could 
conclude that the gradients flattened out from 
$-0.11\,$dex/kpc to $-0.06\,$dex/kpc in about 9 Gyr, or from 
$-0.08\,$dex/kpc to $-0.06\,$dex/kpc in the last 5 Gyr only.
From these data, we can estimate an average flattening rate
of $0.002\,$dex kpc$^{-1}$ Gyr$^{-1}$ (figure 1a)
or a rate of about $0.004\,$dex kpc$^{-1}$ 
Gyr$^{-1}$, considering the last 5 Gyr, suggesting that the 
O/H gradient has not changed more than about 30\% in average 
during the last few Gyr. At earlier times, however, our results 
are consistent with a  steeper rate, although the corresponding 
uncertainties are larger. 

Spatial and time variations of the gradients  are extremely 
important as constraints to chemical evolution models. For
example, recent classical  models predict similar gradients as 
observed, showing some flattening near the outer Galaxy and a 
time steepening (Matteucci 2000), while multiphase models 
(Moll\'a et al. 1997)  generally produce gradientes that were
steeper in the past. Hou et al. (2000) derived a flattening rate 
for O/H of about $0.004$ dex kpc$^{-1}$ Gyr$^{-1}$ for the 
last 10 Gyr, in remarkable agreement with our recent results.
\centerline{\psfig{figure=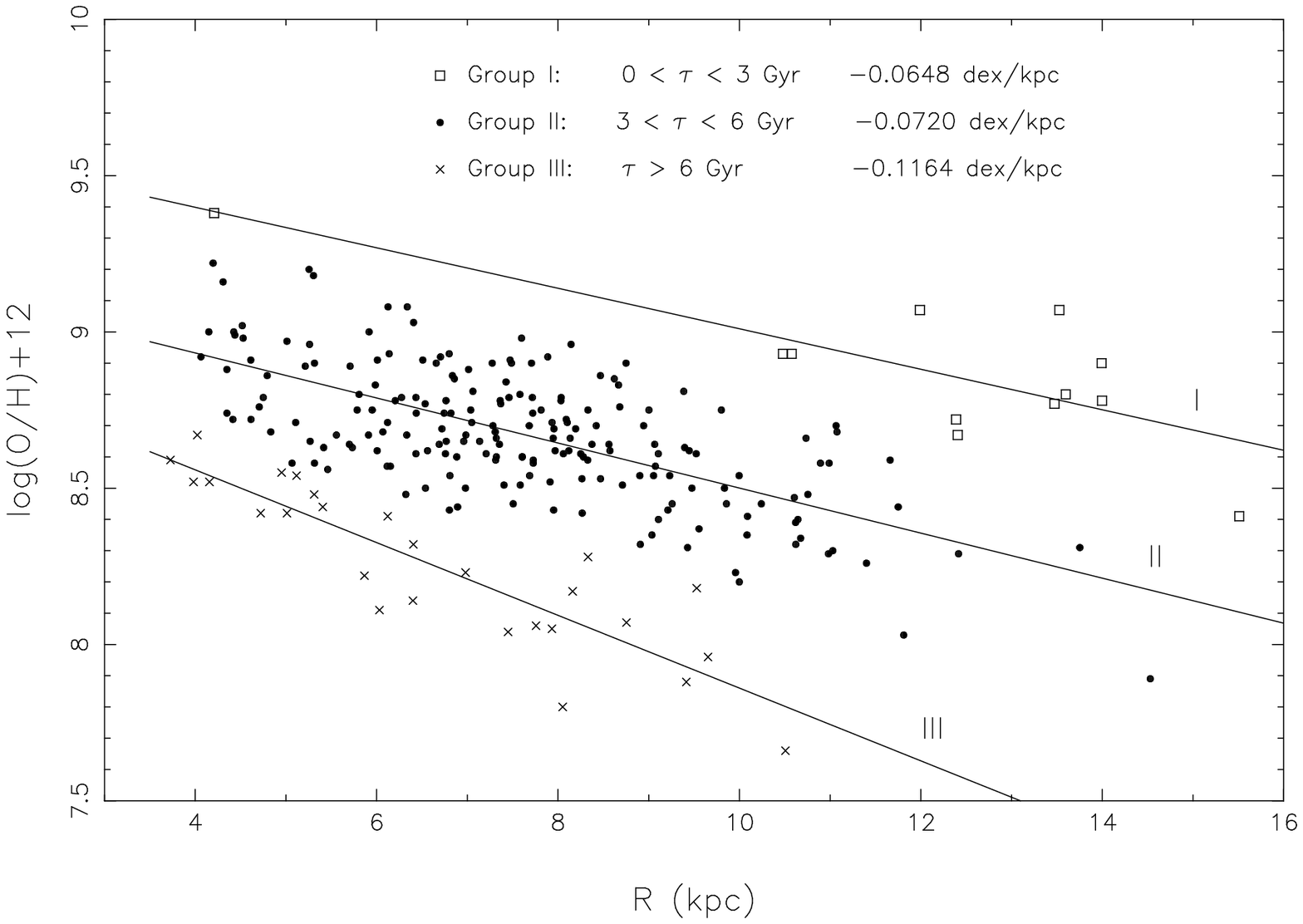,height=9.5 cm,angle=0} }

\centerline{\psfig{figure=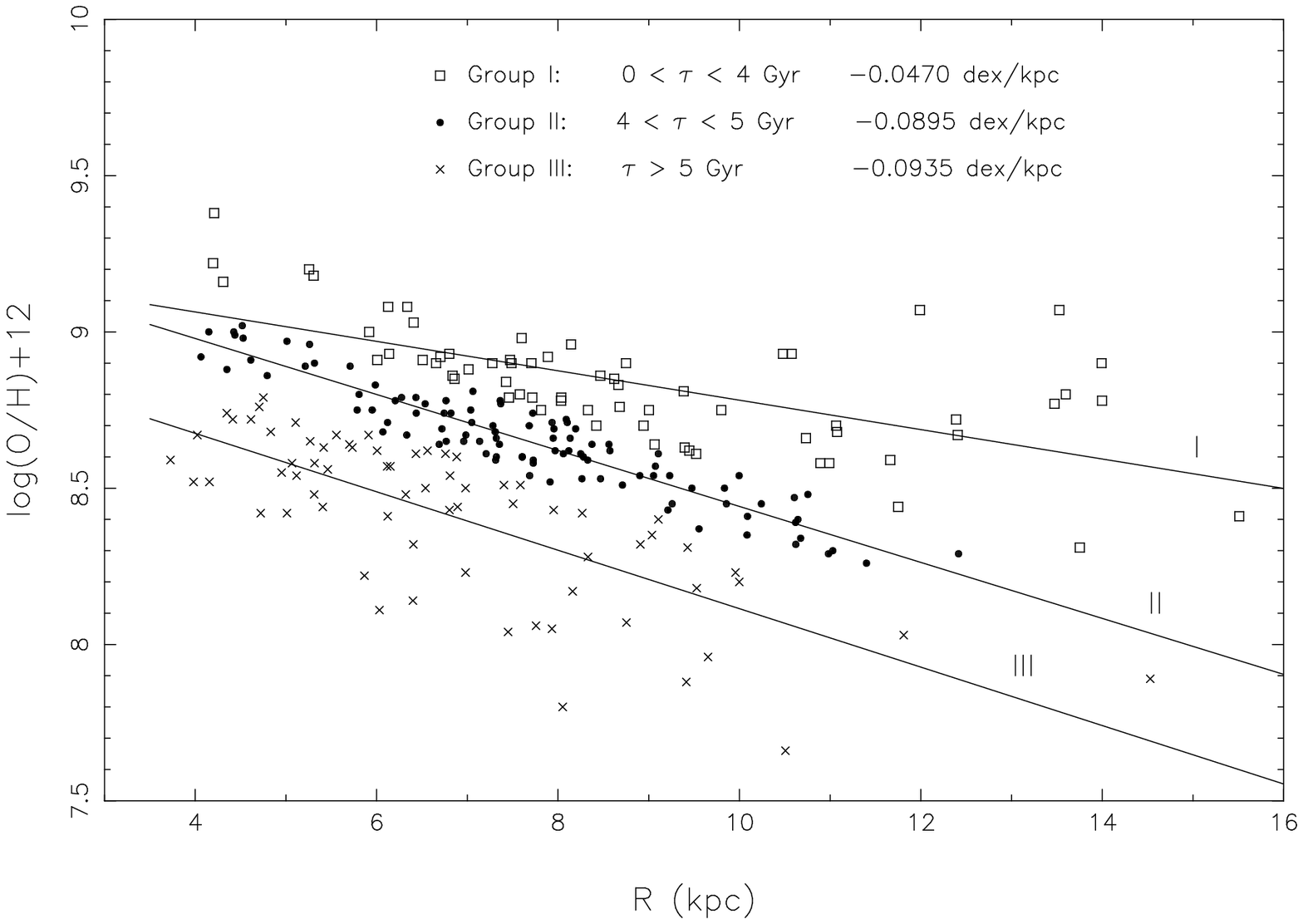,height=9.5 cm,angle=0} }
\bigskip
{Figure 1.\ \ Time variation of the O/H gradient 
    from PN.}
\bigskip
\bigskip
\bigskip\noindent
{\bf 4. \ \ He: PN and Chemical Evolution}
\bigskip\noindent
In order to use PN to investigate the chemical evolution of
\ $^4$He in the Galaxy, it is necessary to take into account 
the He contamination by the progenitor stars. In some previous 
work, this contamination has been introduced in an approximate 
way (Chiappini and Maciel 1994). Maciel (2000b; 2001a) has used 
recent determinations of helium yields in order to estimate the 
$\, ^4$He contamination for a sample of disk planetary nebulae. 
Taking into account some relationships involving the nebular 
abundances, the central star mass and the stellar mass on the 
main sequence, such contamination can be individually determined. 
In practice, the pregalactic He abundance $Y_p$ is not sensitive 
to the PN data, as these nebulae are more metal rich than the 
dwarf galaxies and blue compact galaxies usually considered to 
determine $Y_p$, so that this parameter is essentially fixed by 
these objects. As a consequence, the pregalactic value has been 
taken as a parameter, with the values $Y_p = 0.23$ and $Y_p = 0.24$.

The recent analysis by Maciel (2000b, 2001a)  has shown that the
He/H gradient is essentially flat, irrespective of the 
contamination corrections, that is, the correction procedure 
simply shifts the He abundances downwards, without affecting the 
slope. The amount by which the abundances are reduced depends 
somewhat on the input yields, being typically in the range 
$\Delta$(He/H)$\sim 0.006 - 0.012$. The main conclusion is that 
it is unlikely that any He/H radial gradient could be presently  
detected from PN. These results are used in order to estimate an 
upper limit to the He/H gradient, which is given by 
$\vert d({\rm He/H})/dR \vert  < 0.004 \ {\rm kpc}^{-1}$, 
corresponding to $d\log({\rm He/H})/dR \simeq -0.02$ dex/kpc. 
Therefore, the existence of a He/H radial gradient similar to the 
O/H gradient is extremely unlikely, and any He/H gradient should 
be lower than the O/H gradient by at least a factor of 3. Such a 
conclusion is in good agreement with the small gradients recently 
derived for galactic HII regions by Esteban et al. (1999) and with 
recent chemical evolution models (Chiappini and Matteucci 2000). 

The helium to metals enrichment ratio $\Delta Y/\Delta Z$ is an 
important parameter in the chemical evolution of the Galaxy, and is 
usually determined from HII regions and HII galaxies. Maciel 
(2000b, 2001a) has taken into account all PN in the sample by Maciel 
and Quireza (1999) having metal abundances up to 
$10^6\, {\rm O/H} \simeq 700$, which corresponds approximately to 
the solar value, $\epsilon({\rm O}) = \log {\rm (O/H)} + 12 = 8.83$. 
The relation $Z \simeq 25$ O/H was adopted, where O/H is the 
oxygen abundance by number relative to hydrogen. Results indicate 
that the correction procedure not only reduces the average He 
abundances and the derived $\Delta Y/\Delta Z$ ratio, but also 
decreases the uncertainty of the derived slopes. The average results 
are $2.8 < \Delta Y/\Delta Z < 3.6$ for $Y_p = 0.23$ 
and $2.0 < \Delta Y/\Delta Z < 2.8$ for $Y_p = 0.24$, with a
good agreement with independently derived ratios 
(cf. Maciel 2000b, 2001a).
\bigskip\noindent
{\bf 5. \ \ Oxygen and Iron Abundances}
\bigskip\noindent
The [O/Fe] $\times$ [Fe/H] ratio is one of the basic 
relationships in the study of the chemical evolution of the 
Galaxy, linking the main metallicity indicators and stressing 
the different contribution of Type~II and Type~Ia supernovae. 
Chemical evolution models usually predict different [O/Fe] 
$\times$ [Fe/H] relationships for the galactic disk and bulge, 
reflecting the different rates at which these elements are 
produced in different scenarios, being usually faster in the bulge.
Recently, some discrepancy has been observed between different 
sets of observational data and among theoretical models 
regarding the value of the [O/Fe] abundance ratio at low 
metallicites (see Maciel 2001b,c for references). Generally 
speaking, studies based on the [OI] forbidden line doublet at 
6300 \AA, 6364 \AA\ lead to [O/Fe] $\simeq 0.5$ for [Fe/H] 
$\simeq -2$, showing a plateau in the [O/Fe] ratio at low 
metallicities; on the other hand, oxygen abundances from the 
OI infrared lines and some recent studies based on ultraviolet 
OH bands in metal-poor subdwarfs reach a much higher ratio, 
[O/Fe] $\simeq 1$ at low metallicities, with an essentially 
constant slope of about $-0.30$ to $-0.40$ for the [O/Fe] 
$\times$ [Fe/H] relation.

Some contribution to the understanding of this problem can be 
obtained by the analysis of the radial abundance gradients 
in the galactic disk. Apart from O/H gradients, [Fe/H] 
gradients can also be derived, especially from open cluster 
stars. Maciel (2001b,c) has taken both sets of data into 
account and derived an independent [O/Fe] $\times$ [Fe/H] 
relation appropriate to the galactic disk, roughly at 
metallicities [Fe/H] $\geq -1.5$. 

Assuming that the O/H and [Fe/H] gradients apply essentially 
to the same region in the disk and adopting  average linear 
gradients $d\log ({\rm O/H})/dR \simeq -0.07$ dex/kpc and 
$d$[Fe/H]$/dR \simeq -0.085$ dex/kpc, Maciel (2001b,c) 
obtained a relation given by [O/Fe] $= \alpha + \beta$ [Fe/H],
where $\alpha \simeq 0.098$ and $\beta \simeq -0.176$. This 
can be  seen in Figure 2 (solid line), which shows the 
[Fe/H]  $\times$ $\log$(O/H) + 12 relationship, where 
[Fe/H] $ = \gamma + \delta \, [\log{\rm (O/H)} + 12]$, with 
$\delta = 1/(1 + \beta) \simeq 1.214$ and $-\gamma/\delta = 
\alpha + [\log{\rm(O/H)}_\odot +12]$, or $\gamma \simeq -10.841$. 
Figure 2 is particularly useful for photoionized nebulae such 
as HII regions and PN, for which the [Fe/H] abundance is 
usually difficult to obtain directly. In this case, an average 
relation could be used to estimate the expected [Fe/H] for a 
given O/H ratio. Also, if a determination of [Fe/H] is available, 
this relation could be used to estimate the amount of iron  
condensed in solid grains. The dotted line shows results of 
theoretical models by Matteucci et al. (1999), which predict a 
maximum [O/Fe] $\simeq 0.5$ and the dot-and-dashed line are 
models by Ramaty et al. (2000), which are consistent with higher 
[O/Fe] ratios at low metallicities. The figure also shows some 
representative observational data from several sources.
It can be seen that the gradient data support the lower 
[O/Fe] regime, at least for metallicities larger than 
[Fe/H] $\simeq -1.5$. Extrapolating the solid line towards 
lower metallicities (broken line), we obtain 
an upper limit for the [O/Fe] ratio of 0.4 dex. Therefore, 
these results are consistent with a maximum [O/Fe] $\simeq 0.4$ 
for the galactic disk for metallicities as low as [Fe/H] 
$\simeq -1.5$.

The recent results by Cuisinier et al. (2000) and Costa \& 
Maciel (1999) show that bulge PN have O/H abundances comparable
with their disk counterparts, both regarding the highest 
oxygen abundances attained in the bulge and the metallicity 
range. Since underabundant nebulae are also present, these 
results suggest that the bulge contains a mixed population, 
so that star formation in the bulge  spans a relatively wide 
time interval. 

The PN metallicity distribution can  be compared with the 
stellar abundance distributions, provided we are able to 
convert the measured nebular O/H  abundances into the usual 
[Fe/H]  metallicities. Direct measurements are of limited 
usefulness, in view of the depletion attributed to grain 
formation, but it is 
\centerline{\psfig{figure=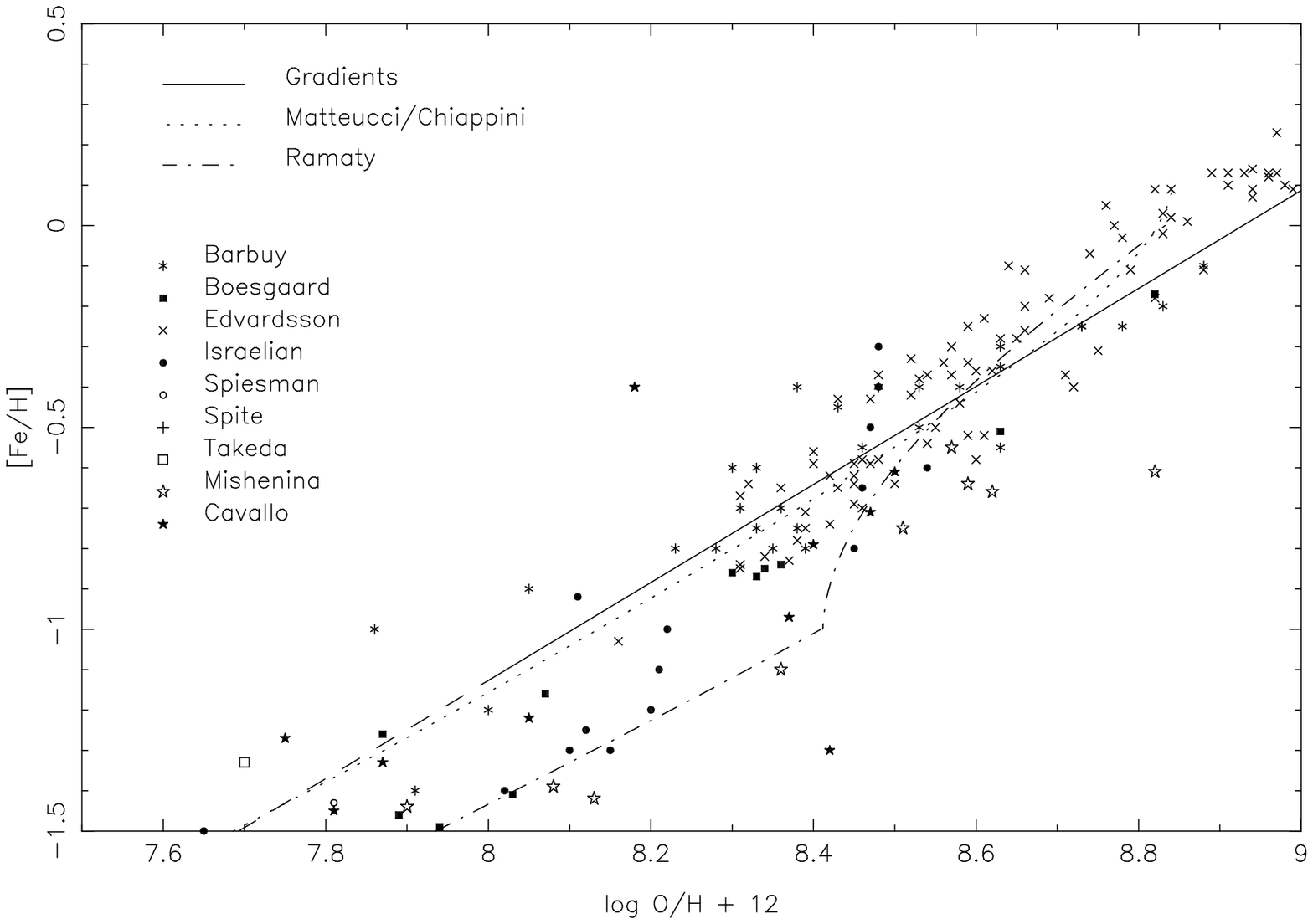,height=7.3 cm,angle=0} }
\medskip
{Figure 2.\ \ The [Fe/H] $\times$ $\log$(O/H) relation 
    in the galactic disk.}
\bigskip
\bigskip
\noindent
possible to convert O/H abundances into 
[Fe/H] metallicities using theoretical [O/Fe] $\times$ [Fe/H] 
relationships, as given by Matteucci et al. (1999).
Maciel (1999) has derived the bulge PN [Fe/H] distribution,
which can be compared with the metallicity distribution 
of bulge K giant stars in Baade's Window (McWilliam \& Rich 1994) 
and with the distribution of bulge Mira variables (Feast \& 
Whitelock 2000). It results that the PN distribution looks 
similar to the K giant distribution if the {\it solar 
neighbourhood}  relation is adopted, but when the {\it bulge} 
relation is taken into account the derived distribution is 
displaced towards lower metallicities by roughly 0.5 dex. 
Maciel (1999) concludes that this discrepancy can be attributed 
to the uncertainties in the adopted [O/Fe] $\times$ [Fe/H] 
relationship for the bulge, which overestimates the 
[O/Fe] enhancement by 0.3 to 0.5 dex. Therefore,
the [O/Fe] $\times$ [Fe/H] relation for the bulge is
closer to the solar neighbourhood relation than implied by the 
theoretical models, which is equivalent of saying that the 
[O/Fe] ratio in the bulge reaches a maximum value of [O/Fe] 
$\sim 0.5$ at metallicities [Fe/H] $\simeq -2.0$.

\smallskip\noindent
{\it Acknowledgements}. We thank FAPESP and CNPq for partial support.
\bigskip
\bigskip\noindent
{\bf References}
\medskip
\ref
{Andrievsky, S.M., Kovtyukh, V.V.,
    Luck, R.E., Barbuy, B., L\'epine, J.R.D., Bersier,
    D., Maciel, W.J., Klochkova, V.G., Panchuk, V.E., \&
    Karpischek, R.U., 2002 A\&A\  (in press)}

\ref
{Caputo, F., Marconi, M., Musella, I., \& Pont, F. 2001, A\&A, 372, 544}

\ref
{Chiappini, C., \& Maciel, W.J. 1994, A\&A,  288, 921}

\ref
{Chiappini, C., \&  Matteucci, F. 2000, in IAU Symp. 198, ed. L. da Silva, 
    M. Spite, \&  J. R. de Medeiros, ASP, 540}

\ref
{Costa, R.D.D., Chiappini, C., Maciel, W.J., \& Freitas Pacheco, J.A. 
    1996  A\&AS, 116, 249}

\ref
{Costa, R.D.D., \&  Maciel, W.J. 1999, Ap\&SS,  265, 327}

\ref
{Cuisinier, F., Maciel, W.J., K\"oppen, J., Acker, A., \&  Stenholm, B.  
    2000 A\&A,  353, 543}

\ref
{Edvardsson, B., Andersen, J., Gustafsson, B., Lambert, D.L., Nissen, 
    P.F., \& Tomkin, J. 1993, A\&A,  275, 101}

\ref
{Esteban, C., Peimbert, M., Torres-Peimbert, S., \&  Garcia-Rojas, J.
    1999, Rev. Mexicana Astron. Astrof.,  35, 65}

\ref
{Feast, M., \& Whitelock, P. 2000, in  Chemical Evolution of the Milky Way, 
    ed. F. Matteucci \& F. Giovannelli (Dordrecht: Kluwer), 229}

\ref
{Ferguson, A.M.N., Gallagher, J.S., \& Wyse, R.F.G. 1998, 
    AJ, 116, 673}

\ref
{Henry, R.B.C., \&  Worthey, G. 1999, PASP, 111, 919}

\ref
{Hou, J.L., Prantzos, N., \&  Boissier, S. 2000, A\&A, 362, 921}

\ref
{Maciel, W.J. 1997, in IAU Symp. 180, ed. H.J. Habing \& H.J.G.L.M. 
    Lamers (Dordrecht: Kluwer), 397}

\ref
{Maciel, W.J. 1999, A\&A, 351, L49}

\ref
{Maciel, W.J. 2000a, in Chemical Evolution of the Milky Way, 
    ed. F. Matteucci \& F. Giovannelli (Dordrecht: Kluwer), 81}

\ref
{Maciel, W.J. 2000b, in IAU Symp. 198, ed. L. da Silva, M. Spite,
    \& J. R. de Medeiros, ASP, 204}

\ref
{Maciel, W.J. 2001a, Ap\&SS\  (in press)}

\ref
{Maciel, W.J. 2001b, New Astron. Rev., 45, 571}

\ref
{Maciel, W.J. 2001c, Rev. Mexicana Astron. Astrof. SC  
    (in press)}

\ref
{Maciel, W.J., \& K\"oppen, J. 1994, A\&A,  282, 436}

\ref
{Maciel, W.J., \& Quireza, C. 1999, A\&A,  345, 629}

\ref
{Matteucci, F. 2000, in Chemical Evolution of the Milky Way, 
    ed. F. Matteucci \& F. Giovannelli (Dordrecht: Kluwer), 3}

\ref
{Matteucci, F., Romano, D., \&  Molaro, P. 1999, A\&A,  341, 458}

\ref
{McWilliam, A., \& Rich, R.M. 1994, ApJ, 91, 749}

\ref
{Moll\'a, M., Ferrini, F., \&  D\'\i az, A.I. 1997, ApJ, 475, 519}

\ref
{Peimbert, M. 1978, in IAU Symp. 76, ed. Y. Terzian 
    (Dordrecht: Reidel), 215}

\ref
{Peimbert, M., \&  Carigi, L. 1998, in ASP Conf. Ser. 147, 
    ed. D. Friedli, M.G. Edmunds, C. Robert \& L. Drissen, 88}

\ref
{Ramaty, R., Scully, S.T., Lingenfelter,R.E., \& Kozlovsky,B. 
   2000 ApJ,  534, 747}

\ref
{Rolleston, W.R.J., Smartt, S.J., Dufton, P.L., \&  Ryans, R.S.I. 
    2000, A\&A, 363, 537}

\ref
{Smartt, S.J. 2000, in Chemical Evolution of the Milky Way, 
    ed. F. Matteucci \& F. Giovannelli (Dordrecht: Kluwer), 323}

\ref
{V\'\i lchez, J.M., \& Esteban, C. 1996, MNRAS, 280, 720}

\bye